\documentclass{article}
\usepackage[margin=1in,heightrounded]{geometry}
\usepackage{hyperref}
\usepackage{multirow}
\usepackage{algorithm}
\usepackage{algorithmic}
\usepackage{subcaption}
\usepackage{graphicx}
\usepackage{amsmath}
\usepackage{amsthm}

\newtheorem{theorem}{Theorem}
\newtheorem{lemma}{Lemma}
\newtheorem{corollary}{Corollary}

\begin{document}
\title{Communication-Efficient Distributed Graph Clustering and Sparsification under Duplication Models
\thanks{An abstract version of this paper was published in the proceedings of CIAC 2023. This work was supported by UNC-Greensboro start-up funds.}}
\author{Chun Jiang Zhu \\
University of North Carolina at Greensboro}
\date{}
%%
% \authorrunning{}
%% First names are abbreviated in the running head.
%% If there are more than two authors, 'et al.' is used.
%%
% \institute{University of North Carolina at Greensboro}
%
\maketitle              % typeset the header of the contribution

\begin{abstract}
In this paper, we consider the problem of clustering graph nodes and sparsifying graph edges over distributed graphs, when graph edges with possibly edge duplicates are observed at physically remote sites. Although edge duplicates across different sites appear to be beneficial at the first glance, in fact they could make the clustering and sparsification more complicated since potentially their processing would need extra computations and communications. We propose the first communication-optimal algorithms for two well-established communication models namely the message passing and the blackboard models. Specifically, given a graph on $n$ nodes with edges observed at $s$ sites, our algorithms achieve communication costs $\tilde{O}(ns)$ and $\tilde{O}(n+s)$ ($\tilde{O}$ hides a polylogarithmic factor), which almost match their lower bounds, $\Omega(ns)$ and $\Omega(n+s)$, in the message passing and the blackboard models respectively. The communication costs are asymptotically the same as those under non-duplication models, under an assumption on edge distribution. Our algorithms can also guarantee clustering quality nearly as good as that of centralizing all edges and then applying any standard clustering algorithm. Moreover, we perform the first investigation of distributed constructions of graph spanners in the blackboard model. We provide almost matching communication lower and upper bounds for both multiplicative and additive spanners. For example, the communication lower bounds of constructing a $(2k-1)$-spanner in the blackboard with and without duplication models are $\Omega(s+n^{1+1/k}\log s)$ and $\Omega(s+n^{1+1/k}\max\{1,s^{-1/2-1/(2k)}\log s\})$ respectively, which almost match the upper bound $\tilde{O}(s+n^{1+1/k})$ for both models.
\end{abstract}


\section{Introduction}
Graph clustering is one of the most fundamental tasks in machine learning.
Given a graph consisting of a node set and an edge set, graph clustering asks to partition graph nodes into clusters such that nodes within the same cluster are ``densely-connected''  by graph edges, while nodes in different clusters are ``loosely-connected''.
Graph clustering on modern large-scale graphs imposes high computational and storage requirements, which are too expensive to obtain from a single machine.
In contrast, distributed computing clusters and server storage are a popular and cheap way to meet the requirements.
Distributed graph clustering has received considerable research interests, \emph{e.g.}, \cite{CSW+16,SZ19,ZZL+19a}. Interestingly, these works show their close relationships with (distributed) graph sparsification.

Graph sparsification is the task of approximating an arbitrary graph by a sparse graph that has a reduced number of edges while approximately preserving certain property.
It is often useful in the design of efficient approximation algorithms, since most algorithms run faster on sparse graphs than the original graphs.
Several notions of graph sparsification have been proposed.
Spectral sparsifiers \cite{ST11} well approximate the spectral property of the original graphs and can be used to approximately solve linear systems over graph Laplacian, and to approximate effective resistances, spectral clustering, and random walk properties \cite{SS11,CSW+16}.
On the other hand, graph spanners are a type of graph sparsifiers that well approximate shortest-path distances in the original graph. A subgraph $H$ of an undirected graph $G$ is called a $k$-spanner of $G$ if the distance between any pair of vertices in $H$ is no larger than $k$ times of that in $G$, and $k$ is called the \emph{stretch} factor.
It is well known that for any $n$-vertex graph, there exists a spanner of stretch $2k-1$ and size (the number of edges) $O(n^{1+1/k})$ \cite{TZ05}.
This is optimal if we believe the Erdos's girth conjecture \cite{Erdos64}.
Many research efforts were then devoted to \emph{additive spanners}, where the distance between any vertex pair is no larger by an additive term $\beta$ instead of a multiplicative factor.
Here the spanner is called a $+\beta$-spanner.
There have been different constructions of +2-, +4-, +6-spanners of size $O(n^{3/2})$, $O(n^{7/5})$, and $O(n^{4/3})$, respectively \cite{BKM+10,Chechik13}.
Spanners have found a wide range of applications in network routing, synchronizers and broadcasting, distance oracles, and preconditioning of linear systems \cite{TZ05,ahmed2020graph}.

In an $n$-vertex distributed graph $G(V,E)$, each of $s$ sites, $S_i$, holds a subset of edges $E_i\subseteq E$ on a common vertex set $V$ and their union is $E=\cup_{i=1}^s E_i$.
We consider two well-established models of communication, the  \emph{message passing} model and \emph{blackboard} model, following the above work.  In the former, there is a communication channel between every site and a distinguished coordinator. Each site can send a message to another site by first sending to the coordinator, who then forwards the message to the destination. In the latter, sites communicate with each other through a shared blackboard such as a broadcast channel.
The models can be further considered in two settings: edge sets of different sites are disjoint (\emph{non-duplication} models) and they can have non-empty intersection (\emph{duplication} models). 
Here the major objective is to minimize the communication cost that is usually measured by the total number of bits communicated.

\begin{figure*}[t]
     \center
     \begin{subfigure}[b]{0.225\textwidth}
        \centering\includegraphics[width=.9\linewidth]{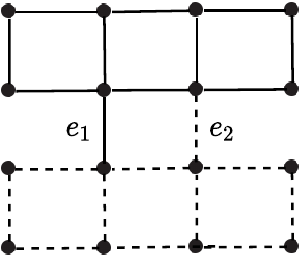}
        \caption{\small A graph $G$ w/o edge duplication}
        \label{fig:wo}
     \end{subfigure}\quad
     \begin{subfigure}[b]{0.225\textwidth}
        \centering\includegraphics[width=.9\linewidth]{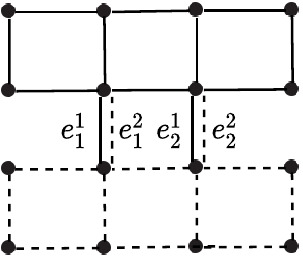}
        \caption{\small A graph $G'$ w/ edge duplication}
        \label{fig:w}
     \end{subfigure}\quad
     \begin{subfigure}[b]{0.225\textwidth}
        \centering\includegraphics[width=.9\linewidth]{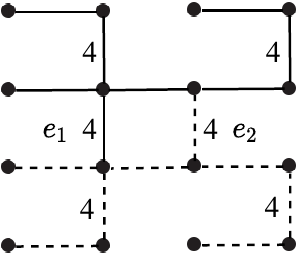}
        \caption{\small The union $H_1\cup H_2=(1+\epsilon)$-$SS(G)$}
        \label{fig:swo}
     \end{subfigure}\quad
     \begin{subfigure}[b]{0.245\textwidth}
        \centering\includegraphics[width=.9\linewidth]{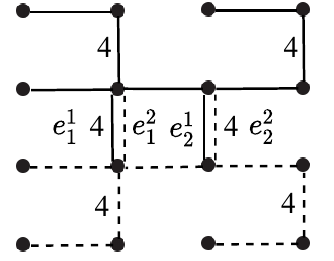}
        \caption{\small The union $H'_1\cup H'_2\not=(1+\epsilon)$-$SS(G')$}
        \label{fig:sw}
     \end{subfigure}
     \caption{\small An illustrating example for challenges in processing edge duplicates across sites. For all subfigures, edge weights are one unless stated explicitly and edges are distributed at two sites: solid edges are in site $S_1$ and dash edges are in $S_2$.  (a): a graph $G$ without edge duplication. The graph $G'$ in (b) is similar to $G$ but edge $e_1$ (and $e_2$) appears in both sites $S_1$ and $S_2$ as $e^1_1$ and $e^2_1$ ($e^1_2$ and $e^2_2$), respectively. (c) shows the decomposability. Each site $S_i$ constructs a spectral sparsifier $H_i$ of its local graph and their union is a spectral sparsifier of $G$. However, the decomposability does not work for $G'$ as in (d). It is unknown how to process the two "duplicates" of $e_1$ and $e_2$, \emph{e.g.}, $e^1_1$ and $e^2_1$ with different weights $4$ and $1$.}
     \label{fig:exam}
     \vskip -0.1in
\end{figure*}

A typical framework of distributed graph clustering is to employ graph sparsification tools to significantly reduce the size of edge sets of different sites while keeping structural properties. \cite{CSW+16} proposed to compute spectral sparsifiers for the graphs at different sites and transmit them to the coordinator. Upon receiving all sparsifiers, the coordinator takes their union and applies a standard clustering algorithm, \emph{e.g.}, \cite{NJW01}. 
%The obtained clustering results have a quality nearly as good as those from a simple method of centralizing all edge sets of different sites and then applying a clustering algorithm. The incurred communication cost is proved to be optimal up to poly-logarithmic factors.
%
However, all the existing methods that follow this framework such as \cite{CSW+16,ZZL+19a} only work in non-duplication models. The assumption that edge sets of different sites are disjoint is crucial to get the \emph{decomposability} of spectral sparsifiers: the union of spectral sparsifiers of subgraphs at different sites is a spectral sparsifier of the distributed graph. Unfortunately, the decomposability does not work in duplication models. When edge sets of different sites have non-empty intersection, it is unclear how to process edge ``duplicates" that are possible to have different edge weights after sparsification. See Figure \ref{fig:exam} for a concrete example.
To the best of our knowledge, none of the existing algorithms can perform distributed graph clustering in the more general duplication models with reasonable theoretical guarantees on both communication cost and clustering quality.
Instead of restoring the decomposability and turning to the framework, our algorithms are built based on the construction of spectral sparsifiers by graph spanners \cite{KX16}. The adaptation of the algorithm to the duplication models need new algorithmic procedures such as weighted graph spanners and uniform sampling. 

Although distributed constructions of graph spanners have been studied in message passing and CONGEST models \cite{CKP+18,FWY20,ZLB21a}, unfortunately they have not been systematically studied in the blackboard model. The blackboard model represents distributed systems with a broadcast channel. It can be viewed as a model for single-hop wireless networks and has received increasingly growing research \cite{CSW+16,dershowitz2021communication,vempala2020communication}. In the second part of this paper, we also investigate the problem of constructing graph spanners under the blackboard with both duplication and non-duplication models and obtain several almost matching communication lower and upper bounds.

\smallskip
{\noindent \bf Our Contributions.}
We perform the first investigation of distributed graph clustering and spectral sparsification under duplication models. We propose communication-optimal (up to polylogarithmic factor) algorithms with communication cost $\tilde{O}(ns)$ and $\tilde{O}(n+s)$ in the message passing and blackboard with duplication models, respectively.
Interestingly, the communication costs are asymptotically the same as the those in the non-duplication models under an assumption on edge distribution: the probability of an edge residing at each of the sites is a known value. This is practical when the popularity or degree of duplication of edges is obtainable.
It is guaranteed that the quality of our clustering results is nearly as good as the simple method of centralizing all edge sets at different sites and then applying a standard clustering algorithm, \emph{e.g.}, \cite{NJW01}.

Furthermore, we study distributed constructions of graph spanners in the blackboard models with and without edge duplication in order to improve our poor understanding on the communication complexity. Table \ref{tab:distributed} summarizes our main findings and Table \ref{tab:distributed-mp} provides the communication complexity in the message passing model \cite{FWY20}. We confirm that the blackboard model is able to significantly reduce the communication complexity compared to the message passing model. Unlike the problem of distributed clustering and spectral sparsification, edge duplication potentially brings more communications for distributed spanner construction problem. See detailed discussions in Section \ref{sec:spanner}.

\begin{table}[t]
	\small
	\center
	\begin{tabular}{|c|c|c|c|}
	\hline
	\multirow{2}{*}{Problem} & \multirow{2}{*}{Upper Bound} & \multicolumn{2}{c|}{Lower Bound} \\ \cline{3-4}
	& & Non-duplication & Duplication \\ \hline 
	
	$(2k-1)$-spanner & $\tilde{O}(s+n^{1+1/k})$ & $\Omega(s+n^{1+1/k}\max\{1,\frac{\log s}{s^{(1+1/k)/2}}\})$ & $\Omega(s+n^{1+1/k}\log s)$ \\ \hline
	
	$+2$ or $3$-spanner & $\tilde{O}(s+n\sqrt{n+s})$ & $\Omega(s+n^{3/2})$ &   $\Omega(s+n^{3/2}\log s)$ \\ \hline
	
	$+k$-spanner & $\tilde{O}(s+n\sqrt{n+s})$ & $\Omega(s+n^{4/3-o(1)})$ & $\Omega(s+n^{4/3-o(1)}\log s)$ \\ \hline 
	\end{tabular}
	\caption{\small Communication complexity of computing graph spanners in the blackboard model, where $n$ is the number of vertices in the input graph and $s$ is the number of sites.}
	\label{tab:distributed}
	\vspace{-0.1in}
\end{table}

\begin{table}[t]
	\small
	\center
	\begin{tabular}{|c|c|c|c|c|}
	\hline
	\multirow{2}{*}{Problem} & \multicolumn{2}{c|}{Upper Bound} & \multicolumn{2}{c|}{Lower Bound} \\ \cline{2-5}
	& Non-duplication & Duplication & Non-duplication & Duplication \\ \hline 
	
	$(2k-1)$-spanner & $\tilde{O}(ks^{1-2/k}n^{1+1/k}+snk)$ & $\tilde{O}(sn^{1+1/k})$ & $\Omega(ks^{1/2-1/(2k)}n^{1+1/k}+sn)$ & $\Omega(sn^{1+1/k})$\\ \hline
	
	$+2$ or $3$-spanner & $\tilde{O}(\sqrt{s}n^{3/2}+sn)$ & $\tilde{O}(sn^{3/2})$ & $\Omega(\sqrt{s}n^{3/2}+sn)$ & $\Omega(sn^{3/2})$ \\ \hline
	
	$+k$-spanner & $\tilde{O}(\sqrt{s/k}n^{3/2}+snk)$ & $\tilde{O}(sn^{3/2})$ & $\Omega(n^{4/3-o(1)}+sn)$ & $\Omega(sn^{4/3-o(1)})$ \\ \hline 
	\end{tabular}
	\caption{\small Communication complexity of computing graph spanners in the message passing model \cite{FWY20}.}
	\label{tab:distributed-mp}
    \vspace{-0.2in}
\end{table}

\smallskip
{\noindent \bf Related Work.}
There have been extensive research on graph clustering in the distributed setting, \emph{e.g.}, \cite{YX15,CSW+16,SZ19,ZZL+19a}.
\cite{YX15} proposed a divide and conquer method for distributed graph clustering.
\cite{CSW+16} used spectral sparsifiers in graph clustering for two distributed communication models to reduce communication cost.
\cite{SZ19} presented a computationally and communication efficient node degree based sampling scheme for distributed graph clustering.
\cite{ZZL+19a} studied distributed dynamic graph clustering based on the monotonicity property of graph sparsification.
However, all these methods assume that there are no edge duplicates across different sites and do not work in the more general duplication setting.
Graph spanners have been studied in the non-distributed model \cite{TZ05,AB16} and a few distributed models \cite{CKP+18,FWY20}.
\cite{CKP+18} studied distributed constructions of pair-wise spanners that approximate distances only for some pairs of vertices in the CONGEST model.
\cite{FWY20} studied distributed construction of a serials of graph spanners in the message passing with and without duplication models. But, there exists no prior work considering such construction in the blackboard model, which has been a widely adopted communication model \cite{BO15,vempala2020communication,dershowitz2021communication}.

\section{Definitions and Notations}
\label{sec:def}
A weighted undirected graph $G(V,E,W)$ consists of a vertex set $V$, an edge set $E$ and a weight function $W$ which assigns a weight $W(e)$ to each edge $e\in E$.
$W$ can be omitted from the presentation if it is clear from the context. Throughout the paper let $n=|V|$ and $m=|E|$ denote the number of vertices and the number of edges in $G$ respectively, and $s$ be the number of remote sites $G$ is observed.
Let $w$ be the maximum edge weight in $G$, \emph{i.e.}, $w=\max_eW(e)$.
We denote by $d_G(u,v)$ the \emph{shortest-path distance} from $u$ to $v$ in $G$.
A $\alpha$-spanner and $+\beta$-spanner for $G$ are a subgraph $H(V,E'\subseteq E)$ of $G$ such that for every $u,v\in V$, $d_H(u,v)\leq \alpha * d_G(u,v)$ and $d_H(u,v)\leq d_G(u,v) + \beta$, respectively.

\section{Distributed Graph Clustering}
\label{sec:clustering}

In this section, we state our distributed graph clustering algorithms in the message passing and blackboard with duplication models. We first discuss challenges introduced by edge duplicates presenting at different sites and then show how we overcome the challenges.

\smallskip
{\bf \noindent Definitions.}
Define the graph \emph{Laplacian} of a graph $G$ as $L = D - A$ where $A$ is the adjacency matrix of $G$ and $D$ is the degree matrix, \emph{i.e.}, a diagonal matrix with the $i$-th diagonal entry equal to the sum over the $i$-th row of $A$. A $(1+\epsilon)$-\emph{spectral sparsifier} of $G$, denoted as $(1+\epsilon)$-$SS(G)$, is a (possibly re-weighted) subgraph $H$ of $G$ such that for every $x\in R^n$, the inequality \[(1-\epsilon)x^TL_Gx\leq x^TL_Hx\leq (1+\epsilon)x^TL_Gx\] holds. Each edge $e$ in $G$ has \emph{resistance} $R(e)=1/W(e)$, and the \emph{effective resistance} between any two vertices $u$ and $v$ in $G$, denoted as $R_G(u,v)$, is defined as the potential difference that has to be applied between them in order to drive one unit of current through the network $G$.

\smallskip
{\bf \noindent Challenges.} Distributed graph clustering algorithms designed for non-duplication models cannot be easily extended to duplication models. We explain the fact using \cite{CSW+16} in the message passing model as an example: every site $S_i$ constructs a spectral sparsifier of its local graph $G_i(V,E_i)$ as a synopsis $H_i$ and then transmits $H_i$, instead of $G_i$, to the coordinator. Upon receiving $H_i$ from all sites, the coordinator takes their union, $H=\cup_{i=1}^sH_i$ as the constructed structure.  The algorithm is based on the decomposability property of spectral sparsifiers. 
To see this, for every $i\in[1,s]$, by definition of spectral sparsifiers,  we have for every vector $x\in R^n$,
$(1-\epsilon)x^TL_{G_i}x \leq x^TL_{H_i}x \leq (1+\epsilon)x^TL_{G_i}x.$
Summing all inequalities for $i\in[1,s]$, we get that
\begin{equation*}
    (1-\epsilon)\sum_{i\in[1,s]}x^TL_{G_i}x  \leq \sum_{i\in[1,s]}x^TL_{H_i}x 
     \leq (1+\epsilon)\sum_{i\in[1,s]}x^TL_{G_i}x.
\end{equation*}
In the non-duplication model, it is easy to check that $\sum_{i=1}^sL_{G_i}=L_G$ by the definition of Laplacian matrix. Then the above inequality is equivalent to
\begin{equation}
(1-\epsilon)x^TL_{G}x \leq x^TL_{H}x \leq (1+\epsilon)x^TL_{G}x,
\label{eq:res}
\end{equation}
which concludes that $H$ is a $(1+\epsilon)$-spectral sparsifier of $G$.
Under the duplication model, however, it is clear that $\sum_{i=1}^sL_{G_i}\not=L_G$ and thus Inequality (\ref{eq:res}) does not hold any longer. In other words, the structure $H$ constructed using the same principle is not a spectral sparsifier of $G$. See Figure \ref{fig:exam} for an illustrating example.

\smallskip
{\bf \noindent Proposed Method.} 
Restoring the decomposability of spectral sparsifiers in the duplication models appears to be quite challenging. We avoid it by asking every site cooperates to construct a spectral sparsifier of the distributed graph in the coordinator, who can then get clustering results by any standard clustering algorithm. A standard method of computing spectral sparsifiers \cite{SS11} is to sample each edge in the input graph with a probability proportional to its effective resistance and then include the sampled edges (after appropriate weight rescaling) into the sparsifier. But, when there are duplicated edges across different sites, an edge $(u,v)$ may get sampled more than once at different sites, thereby resulting in multiple edges of possibly different weights between $u$ and $v$, \emph{e.g.}, edges $e^1_1$ and $e^2_1$ in Figure \ref{fig:exam}. It is unclear how to process these edges to guarantee the resulting structure is always a spectral sparsifier. As in Figure \ref{fig:exam}, simply taking union by summing edge weights does not produce a valid spectral sparsifier. 

Instead of using the classic sampling method, we propose to make use of the fact that spectral sparsifiers can be constructed by graph spanners \cite{KX16} to compute spectral sparsifiers in the coordinator. The connection between spectral sparsifiers and graph spanners allows us to convert spectral sparsification to graph spanner construction and uniform sampling under duplication models. 
In the followings, we first introduce the algorithm of \cite{KX16} and then discuss how to adapt the algorithm in the message-passing and blackboard under duplication models.

\smallskip
{\noindent \bf The algorithm of \cite{KX16}.} 
Given a weighted graph, their algorithm first determines a set of edges that has small effective resistance through graph spanners. Specifically, it constructs a $t$-bundle $\log n$-spanner $J=J_1\cup J_2\cup \cdots \cup J_t$, that is, a sequence of $\log n$-spanners $J_i$ for each graph $G_i=G-\cup_{j=1}^{i-1}J_j$ with $1\leq i\leq t=O(\epsilon^{-2}\log n)$. Intuitively, it peels off a spanner $J_i$ from the graph $G_i$ to get $G_{i+1}$ before computing the next spanner $J_{i+1}$, \emph{i.e.}, $J_1$ is a spanner of $G$, $J_2$ is a spanner of $G-J_1$, \emph{etc}. The $t$-bundle spanner guarantees that each \emph{non-spanner} edge (edge not in the spanner) has $t$ edge-disjoint paths between its endpoints in the spanner (and thus in $G$), serving as a certificate for its small effective resistance. The algorithm then uniformly samples each non-spanner edge with a fixed constant probability, \emph{e.g.}, $0.25$ and scales the weight of each sampled edge proportionally, \emph{e.g.}, by $4$ to preserve the edge’s expectation.  By the matrix concentration bounds, it is guaranteed that the spanner together with the sampled non-spanner edges are a moderately sparse spectral sparsifier, in which the number of edges has been reduced by a constant factor. The desirable spectral sparsifier can be obtained by repeating the process until we get a sufficient sparsity, which happens after logarithmic iterations.

\smallskip
{\noindent \bf Weighted Graph Spanners.}
An important building block in \cite{KX16} is the construction of graph spanners of stretch factor $\log n$, which can be used to construct the $t$-bundle $\log n$-spanner. Unfortunately, there is no algorithm that can generate such a spanner under the duplication models. \cite{FWY20} developed an algorithm for constructing $(2k-1)$-spanners in unweighted graphs under the message passing with duplication model through the implementation of the greedy algorithm \cite{ADD+93}. But the algorithm does not work in weighted graphs, where the greedy algorithm would need to process the edges in nondecreasing order of their weights. This seems to be a notable obstacle in both the message passing model and the blackboard model.

In this paper, we first propose an algorithm for constructing $(4k-2)$-spanners in weighted graphs under the message passing with duplication model. We are able to overcome the challenge in weighted graphs at the expense of a larger stretch factor $4k-2$. However, this is sufficient for the construction of $\log n$-spanners in weighted graphs by setting the parameter $k=O(\log n)$.

Specifically, we divide the range of edge weights $[1,w]$ into logarithmic intervals, where the maximum edge weight $w$ is assumed to be polynomial in $n$ \footnote{This is a common and practical assumption for modern graphs.}. Then we process edges in each logarithmic scale $[2^{i-1},2^{i})$, where $1 \leq i \leq log_2(nw)$, as follows.
Each site $S_j$ in order decides which of its edge $e\in E_j$ of weight in $[2^{i-1},2^{i})$ to include into the current spanner $H$. If including the edge $e$ results in a cycle of at most $2k-1$ edges, then the shortest distance between $e$'s endpoints in the current spanner is guaranteed to be less than $(4k-2)W(e)$ (see our proof below). Thus the edge can be discarded. Otherwise, we update the current spanner $H$ by including $e$. After completing processing of $E_j$, $S_j$ forwards the possibly updated spanner $H$ to the next site. The algorithm is summarized in Algorithm (Alg.) \ref{alg:spanner}.

\begin{algorithm}[t]
\small
\caption{$Spanner(G,k)$: $(4k-2)$-spanners under duplication models}
\renewcommand{\algorithmicrequire}{\textbf{Input:}}
\renewcommand{\algorithmicensure}{\textbf{Output:}}
\begin{algorithmic}[1]
\REQUIRE Graph $G(V,E,W)$ and a parameter $k>1$
\ENSURE Spanner $H$
\STATE $H\leftarrow \emptyset$
\FOR{$i\in [1,log_2(nw)]$}
    \FOR{each site $S_j$} 
        \STATE Wait for $H$ from site $S_{j-1}$
        \FOR{each edge $e\in E_j$ of weight in $[2^{i-1},2^i)$} 
            \IF{$(V,H\cup \{e\})$ does not contain a cycle of $\leq 2k$ edges} 
                \STATE $H\leftarrow H \cup \{e\}$
            \ENDIF
        \ENDFOR
        \STATE  Transmit $H$ to the next site $S_{j+1}$ \label{alg:spanner:transmit}
    \ENDFOR
\ENDFOR
\RETURN $H$;
\end{algorithmic}
\label{alg:spanner}
\end{algorithm}

% \begin{algorithm}[t]
% \small
% \caption{$(2k-1)$-spanners in the blackboard model}
% \renewcommand{\algorithmicrequire}{\textbf{Input:}}
% \renewcommand{\algorithmicensure}{\textbf{Output:}}
% \begin{algorithmic}[1]
% \REQUIRE Graph $G(V,E)$ and a parameter $k>1$
% \ENSURE Spanner $H$
% \STATE $H\leftarrow \emptyset$\;
% \FOR{each site $S_i$} 
%     \FOR{each edge $e=(u,v)\in E_i$} 
%         \IF{$d_H(u,v)>(2k-1)d_G(u,v)$}
%             \STATE $H=H\cup \{e\}$; Transmit $\{e\}$ to the blackboard\;
%         \ENDIF
%     \ENDFOR
% \ENDFOR
% \RETURN $H$;
% \end{algorithmic}
% \label{alg:bl}
% \end{algorithm}

\begin{theorem}
\label{thm:spanner-mp}
Given a weighted graph and a parameter $k>1$, Alg. \ref{alg:spanner} constructs a $(4k-2)$-spanner using communication cost $\tilde{O}(sn^{1+1/k})$ in the message passing with or without duplication model.
\end{theorem}

\proof
We first prove that the stretch factor is $4k-2$. For each edge $(u, v) \in E$, if $(u, v) \not\in H$, it must be that including the edge $(u, v)$ would close a cycle of length $\leq 2k$. That is, there exists a path $P$ of $\leq 2k-1$ edges between $u$ and $v$ in $H$. Since we process edges in logarithmic scale, the edge weights in $P$ cannot be larger than $2W(u,v)$. Thus the path length of $P$ is at most $(4k-2)W(e)$. Therefore, the output $H$ is a $(4k-2)$-spanner. 

We then prove the communication cost. By construction, the output graph $H$ has girth (the minimum number of edges in a cycle contained in the graph) larger than $2k$. It is well known that a graph with girth larger than $2k$ have $O(n^{1+1/k})$ edges \cite{ADD+93}. Then $H$ always has $O(n^{1+1/k})$ edges throughout the processing of each logarithmic interval. Thus the total communication cost is $\tilde{O}(sn^{1+1/k})$. The algorithm works for both with and without duplication settings, which do not affect the communication complexity.
\qed

Alg. \ref{alg:spanner} can be extended to the blackboard model with the following modification: In Line \ref{alg:spanner:transmit}, if site $S_j$ does change $H$ by adding some edge(s), it transmits the updated spanner $H$ to the blackboard, instead of the next site; otherwise, it sends a special marker of one bit to the blackboard to indicate that it has completed the processing. The results are summarized in Theorem \ref{thm:spanner-bl}. In Section \ref{sec:spanner}, we will show that the communication cost can be reduced to $2k-1$ in unweighted graphs.

\begin{theorem}
\label{thm:spanner-bl}
The communication complexity of constructing a $(4k-2)$-spanner in weighted graphs under the blackboard with or without duplication model is $\tilde{O}(s+n^{1+1/k})$. In unweighted graph, the stretch factor can be reduced to $2k-1$.
\end{theorem}

\smallskip
{\noindent \bf Constructing $t$-bundle $\log n$-spanner.}
Recall that a $t$-bundle $\log n$-spanner $J=J_1\cup J_2\cup \cdots \cup J_t$, where $J_i$ is a $\log n$-spanner for graph $G_i=G-\cup_{j=1}^{i-1}J_j$, for $1\leq i\leq t$.
When $i=1$, $G_1=G$ is a distributed graph with each site $S_j$ having edge set $E_j$. We can use Alg. \ref{alg:spanner} with $k=(2+\log n)/4$ to compute a $\log n$-spanner $J_1$ of $G_1$. For $2\leq i\leq t$, $G_j=G_{j-1}-J_j$ is again a distributed graph: each site $S_j$ knows which of its edges $E_j$ was included in $J_1, J_2, \cdots, J_{i-1}$ and those edges are excluded from its edge set $E_j-J_1-J_2-\cdots-J_{i-1}$.
Therefore, the construction of a $t$-bundle $\log n$-spanner invokes Alg. \ref{alg:spanner} for $t$ times. Because of $t=O(\epsilon^{-2}\log n)$ and Theorems \ref{thm:spanner-mp} and \ref{thm:spanner-bl}, the total communication costs in the message passing and blackboard with duplication models are $\tilde{O}(sn)$ and $\tilde{O}(s+n)$, respectively.

\smallskip
{\noindent \bf Uniform Sampling.}
After the spanner construction, the algorithm of \cite{KX16} then uniformly samples each non-spanner edge with a fixed probability, \emph{e.g.}, $0.25$ and scales the weight of each sampled edge proportionally, \emph{e.g.}, by $4$.
We observe that sampling with a fixed probability is much more friendly to edge duplicates as compared to sampling with a varied probability used in traditional methods such as \cite{FHH+11}. For example in Figure \ref{fig:exam}, if the duplicates $e_1^1$ and $e_1^2$ of $e_1$ are both sampled (under a fixed probability $0.25$), they still have the same weight $4W(e_1)$ and are edge duplicates again in the next iteration. If one of them, say $e_1^1$, is not sampled, it is removed from the (local) graph at site $S_1$ and will not formulate duplicates with $e_1^2$ at site $S_2$. In contrast, non-uniform sampling could result in sampled edges of rather different weights, which may not be even considered as duplicates.
However, uniform sampling under duplication models is still very challenging: if a fixed probability is used for every edge, an edge with $d$ duplicates across different sites is processed/sampled for $d$ times, each at one of the $d$ sites, and thus has a higher probability being sampled than another edge with smaller duplicates. This results in a non-uniform sampling.

To achieve the uniform sampling, we suppose that the probability of an edge $e$ residing at each of the sites is a known value $r_e$. If we set the probability of random sampling at each site as $p_e$, then the probability that the edge is not sampled at each site is $1-p_e*r_e$. It can be derived that the probability that $e$ is sampled by \emph{at least} one site is $p=1-(1-p_e*r_e)^s$. Since the values of $r_e$ and $s$ are known, we can tune the value of $p_e$ to get the expected sampling probability $p=0.25$. At some site, if $e$ is sampled and added to $H$, we update its presenting probability as $p_e*r_e$, which will be used in the next iteration. Otherwise (if $e$ is not sampled), it is discarded and will not participate in the next iteration. See the details in Algorithms \ref{alg:light} and \ref{alg:ss}.

\begin{algorithm}[t]
\small
\caption{\emph{Light-SS} under duplication models}
\renewcommand{\algorithmicrequire}{\textbf{Input:}}
\renewcommand{\algorithmicensure}{\textbf{Output:}}
\begin{algorithmic}[1]
\REQUIRE $G(V,E),\epsilon\in(0,1)$, and probability $r_e$ for each edge $e$
\ENSURE $H$ with updated $r'_e$ for each edge $e\in H$

\STATE $G_1\leftarrow G$; $J\leftarrow \emptyset$\;
\FOR{$i\in [1,24\log^2n/\epsilon^2]$}
    \STATE $J_i\leftarrow Spanner(G_i,(2+\log n)/4)$
    \STATE $G_{i+1}\leftarrow G_i-J_i$
\ENDFOR
\STATE $H\leftarrow J$; $r'_e\leftarrow r_e$\;
\FOR{each site $S_i$}  \label{light:samplestart}
    \FOR{each edge $e\in E_i-J$}
        \STATE Sample the edge $e$ with probability $p_e$ such that $1-(1-p_e*r_e)^s=0.25$; if $e$ is sampled, adds $e$ to $H$ with a new weight $4W(e)$ and set $r'_e$ to $p_e*r_e$\; \label{light:sampleend}
    \ENDFOR
    \IF{it is the last iteration of the for-loop in Line \ref{ss:for} of Alg. \ref{alg:ss}}
        \STATE Transmit the sampled edges to the coordinator\;
    \ENDIF
\ENDFOR
\RETURN $H$;
\end{algorithmic}
\label{alg:light}
\end{algorithm}

\begin{algorithm}[t]
\small
\caption{$(1+\epsilon)$-\emph{SS} under duplication models}
\renewcommand{\algorithmicrequire}{\textbf{Input:}}
\renewcommand{\algorithmicensure}{\textbf{Output:}}
\begin{algorithmic}[1]
\REQUIRE $G(V,E)$, probability $r_e$ for each edge $e$, and parameters $\epsilon\in(0,1)$ and $\rho>1$
\ENSURE $H$
\STATE $G_0\leftarrow G$\;
\FOR{$i\in [1,\lceil\log\rho\rceil]$} \label{ss:for}
    \STATE $G_i\leftarrow$ \emph{Light-SS}$(G_{i-1},\epsilon/\lceil\log\rho\rceil, r_e)$\; \label{ss:call}
\ENDFOR
%All sites transmit edges in $G_{\lceil\log\rho\rceil}$ to the coordinator; if there are edge duplicates, keep only one edge\;
\STATE $H\leftarrow G_{\lceil\log\rho\rceil}$\; \COMMENT{$H$ is already transmitted to and known by the coordinator}
\RETURN $H$;
\end{algorithmic}
\label{alg:ss}
\end{algorithm}

The main algorithm, Alg. \ref{alg:ss} computes $(1+\epsilon)$-spectral sparsifier in $\lceil\log\rho\rceil$ iterations of \emph{Light-SS}, where $\rho$ is a sparsification parameter. The communication cost of \emph{Light-SS} is composed of the cost for the bundle spanner construction and the cost for non-spanner edge sampling. If the sampled edges are transmitted to the coordinator, the communication cost $\tilde{O}(m)$ could be prohibitively large. To see this, the number of edges in the output $G_i$ after each iteration is only reduced by a constant factor because the uniform sampling removes $3/4$ of the non-spanner edges in expectation. 
To improve the communication cost, we keep sampled edges in each iteration at local sites and do not transmit them to the coordinator except for the very last iteration. Then similar to the input graph $G$, the output $G_i$ for each iteration $i\in [1,\lceil\log\rho\rceil-1]$ are also a distributed graph with possible edge duplication. Edge duplicates come from two sources: either the edge is included into the bundle spanner, or the edge is sampled by more than one site. 
In this way, the communication cost of \emph{Light-SS} (except for the last iteration) contains only the cost of constructing the bundle spanner. In the last iteration, the number of sampled edges must be small $\tilde{O}(n)$, which is also the communication cost of their transmission. Therefore, the communication costs of Alg. \ref{alg:ss} in the message passing and blackboard under duplication models are $\tilde{O}(ns)$ and $\tilde{O}(n+s)$, respectively. Putting all together, our results for distributed spectral sparsification under duplication models are summarized in Theorem \ref{thm:distributed} with its formal proof deferred to Appendix \ref{adx:ss}. 

\begin{theorem}[Spectral Sparsification under Duplication Models]
\label{thm:distributed}
For a distributed graph $G$ and parameters $\epsilon\in (0,1)$ and $\rho=O(\log n)$, Alg. \ref{alg:ss} can construct a $(1+\epsilon)$-spectral sparsifier for $G$ of expected size $\tilde{O}(n)$ using communication cost $\tilde{O}(ns)$ and $\tilde{O}(n+s)$ in the message passing and blackboard with duplication models respectively, with probability at least $1-n^{-c}$ for constant $c$.
\end{theorem}

% Here we relax the assumption on the data distribution.
% Each site samples non-spanner edge with a fixed probability $0.25$ and transmitted the sampled edges to the coordinator. In the blackboard model, each site, once seeing a sampled edge that is not sampled by it, sends a message to the coordinator to remove that edge. In this way, each edge is sampled with the exact probability $0.25$. Because in the worst case, all the non-spanner edges are transmitted, the communication cost is $O(n\log^3n\log^3\rho/\epsilon^2+m/\rho)$

\smallskip
{\noindent \bf Clustering in the Sparsifier.}
After obtaining the spectral sparsifier of the distributed graph, the coordinator applies a standard clustering algorithm such as \cite{NJW01} in the sparsifier to get the clustering results. We can guarantee a clustering quality nearly as good as the simple method of centralizing all graph edges and then performing a clustering algorithm. Before formally stating the results, we define a few notations.
% This also immediately gives distributed clustering algorithms that achieve optimal communication costs. 

For every node set $S$ in a graph $G$, let its \emph{volume} and \emph{conductance} be $vol_G(S)=\sum_{u\in S,v\in V}W(u,v)$ and $\phi_G(S)=(\sum_{u\in S,v\in V-S}W(u,v))/vol_G(S)$, respectively.
Intuitively, a small value of conductance $\phi(S)$ implies that nodes in $S$ are likely to form a cluster.
%Intuitively, a small value of conductance $\phi(S)$ means that nodes in $S$ have few connections with the rest nodes in $V$, and thus implies that nodes in $S$ formulate a cluster.
A collection of subsets $A_1,\cdots,A_k$ of nodes is called a \emph{(k-way) partition} of $G$ if (1) $A_i\cap A_j=\emptyset$ for $1\leq i\not=j\leq k$; and (2) $\cup_{i=1}^kA_i=V$.
The \emph{k-way expansion constant} is defined as $\rho(k)=\min_{partition A_1,\cdots,A_k}\max_{i\in[1,k]}\phi(A_i)$.
A lower bound on $\Upsilon_G(k)=\lambda_{k+1}/\rho(k)$ implies that $G$ has exactly $k$ well-defined clusters \cite{PSZ15}, where $\lambda_{k+1}$ is the $k+1$ smallest eigenvalue of the normalized Laplacian matrix.
For any two sets $X$ and $Y$, their symmetric difference is defined as $X\Delta Y=(X-Y)\cup(Y-X)$.

\begin{theorem}
\label{thm:clustering}
For a distributed graph $G$ with $\Upsilon_G(k)=\Omega(k^3)$ and an optimal partition $P_1,\cdots,P_k$ achieving $\rho(k)$ for some positive integer $k$, there exists an algorithm that can output partition $A_1,\cdots,A_k$ at the coordinator such that for every $i\in [1,k]$, $vol(A_i\Delta P_i)=O(k^3\Upsilon^{-1}vol(P_i))$ holds with probability at least $1-n^{-c}$ for constant $c$. The communication costs in the message passing and blackboard with duplication models are $\tilde{O}(ns)$ and $\tilde{O}(n+s)$, respectively.
\end{theorem}

To the best of our knowledge, this is the first algorithm for performing distributed graph clustering in the message passing and blackboard with edge duplication models. Remarkably, we can show that the communication costs are \emph{optimal}, almost matching the communication lower bounds $\Omega(ns)$ and $\Omega(n+s)$, respectively. It is interesting to see that the communication costs incurred under duplication models are asymptotically the same as those under non-duplication models. In other words, edge duplication does not incur more communications in the graph clustering task, unlike other problems such as graph spanner construction as we will show in Section \ref{sec:spanner}.
Although we make an assumption on the edge distribution probability, we conjecture that when the assumption is relaxed, \emph{i.e.}, graph edges are presenting at different sites arbitrarily, the communication upper bounds remain the same in duplication models. We leave the study as an important future work.

% \begin{conjecture}
% For an $n$-vertex graph $G$ distributed at $s$ sites \textbf{arbitrarily} and a parameter $\epsilon>0$, there exists an algorithm that constructs a $(1+\epsilon)$-spectral sparsfier of $G$ using communication cost $\tilde{O}(ns)$ and $\tilde{O}(n+s)$ in the message passing and blackboard models with edge duplication, respectively.
% \end{conjecture}

% Note that the constructed spectral sparsifier $H$ is a distributed graph with edges appearing at different sites, similar to the input graph $G$. We do not transmit the sampled edges to the coordinator except for the last iteration. If an edge is sampled by more than one sites, then there are edge duplicates at each of these sites.

% This is the \emph{first} algorithm for constructing spectral sparsifiers in the message passing with duplication model. Remarkably, it uses a communication cost $\tilde{O}(ns)$, the same as the without duplication case. We can prove that it is communication \emph{optimal} by showing a matching lower bound of $\Omega(ns)$. However, our results are based on the assumption that the probability of each edge residing at a site is known. We will leave the problem whether we can still achieve the same communication complexity when removing the assumption as a future work.

\section{Spanner Constructions in the Blackboard Model}
\label{sec:spanner}

In this section, we study distributed constructions of graph spanners in the blackboard models with and without edge duplication. This, unfortunately, has not been investigated by prior work yet. We prove several interesting communication upper and lower bounds for typical graph spanners as summarized in Table \ref{tab:distributed}. Due to limit of space, we cannot enumerate every result in Table \ref{tab:distributed}. Hence, here we only describe the general $(2k-1)$-spanners and move the additive spanners to the Appendix. We start with the duplication model, followed by the non-duplication model. The lower bounds obtained in Theorems \ref{thm:spanner-lb} and \ref{thm:spanner-lb-non} hold in both weighted and unweighted graphs and the rest results are on unweighted graphs.

% In this following, we present our results for constructions of different types of graph spanners in the blackboard with and without duplication models. All missing algorithms and their proof are deferred to the Appendix.
% Here we describe techniques for developing communication upper and lower bounds in constructing $(2k-1)$-spanners. 

\smallskip
{\noindent \bf Duplication Model.}
In Section \ref{sec:clustering}, we have provided the communication upper bound, $\tilde{O}(s+n^{1+1/k})$, of constructing $(2k-1)$-spanners in unweighted graphs in Theorem \ref{thm:spanner-bl}. We now show that the communication lower bound is $\Omega(s+n^{1+1/k}\log s)$.

\begin{theorem}
\label{thm:spanner-lb}
The communication lower bound of constructing a $(2k-1)$-spanner in the blackboard with duplication model is $\Omega(s+n^{1+1/k}\log s)$.
\end{theorem}

\proof
To prove this, we target a more general statement that works for every spanner.

\begin{lemma}
\label{lemma:lb}
Suppose there exists an $n$-vertex graph $F$ of size $f(n)$ such that $F$ is the only spanner of itself or no proper subgraph $F'$ of $F$ is a spanner. Then the communication complexity of computing a spanner in the blackboard with duplication model is $\Omega(s+f(n)\log s)$ bits. 
\end{lemma}

\proof
Our proof is based on the reduction from the Multiparty Set-Disjointness problem ($DISJ_{m,s}$) to graph spanner computation. In $DISJ_{m,s}$, $s$ players receive inputs $X_1,X_2,\cdots,X_s\subseteq \{1,\cdots,m\}$ and their goal is to determine whether or not $\cap_{i=1}^sX_i=\emptyset$. 
Now we construct a distributed graph $G$ from the graph $F$ and an instance of $DISJ_{f(n),s}$ as follows. We add edge $e_j$ in $F$ to site $i$ if $j\not\in X_i$ for $1\leq j\leq f(n)$. If the coordinator outputs $F$ as the spanner, we report $\cap_{i=1}^sX_i=\emptyset$; otherwise we report $\cap_{i=1}^sX_i\not=\emptyset$. It can be seen that the coordinator outputs $F$ iff all its edges appear at some site, which is the case $\cap_{i=1}^sX_i=\emptyset$. Finally, according to the communication lower bound of $DISJ_{m,s}$ in the blackboard model \cite{BO15}, $\Omega(s+m\log s)$, the communication complexity of computing a spanner is $\Omega(s+f(n)\log s)$.
\qed

%$H$ is a $(2k-1)$-spanner of $G$
For the lower bound of $(2k-1)$-spanners, the Erdos's girth conjecture states that there exists a family of graphs $F$ of girth $2k+1$ and size $\Omega(n^{1+1/k})$ \cite{Erdos64}. This implies that there exists only one $(2k-1)$-spanner of $F$, that is $F$ itself. It is because the deletion of any edge in $F$ would result in that the distance between the endpoints of the edge becomes at least $2k$. Then by Lemma \ref{lemma:lb}, we get the lower bound $\Omega(s+n^{1+1/k}\log s)$.
\qed

{\noindent \bf Non-Duplication Model.}
In the non-duplication model, we prove a lower bound via a reduction from the lower bound for the duplication model.
% This is inspired by a similar reduction in the message passing models \cite{FWY20}. 

\begin{theorem}
\label{thm:spanner-lb-non}
The communication complexity of constructing a $(2k-1)$-spanner in the blackboard without duplication model is $\Omega(s+n^{1+1/k}\max\{1,s^{-1/2-1/(2k)}\log s\})$.
\end{theorem}

\begin{figure}[t]
     \centering\includegraphics[width=.8\linewidth]{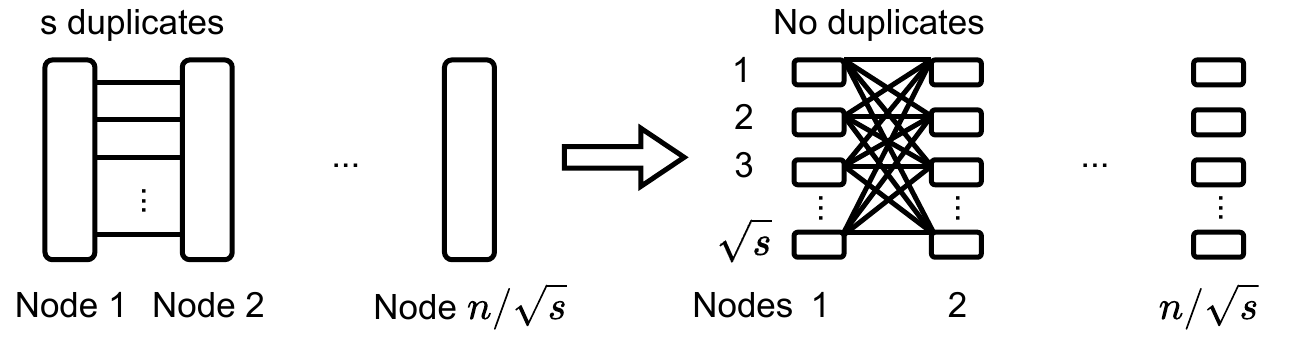}
     \caption{\small Converting a graph with duplication on $s$ sites and $n/\sqrt{s}$ vertices into a graph without duplication on $s$ sites and $n$ vertices}
     \label{fig:lb-exam}
\vspace{-0.2in}
\end{figure}

\proof
We can construct an instance of the $(2k-1)$-spanner problem without duplication on $s$ sites and $n$ vertices from an instance of the $(2k-1)$-spanner problem with duplication on $s$ sites and $n/\sqrt{s}$ vertices. Specifically, we construct a graph $G'$ with no duplication by replacing each vertex $v$ by a set of vertices $S_v$ of size $\sqrt{s}$. Since there are at most $s$ copies of an edge $(u,v)$ in the original graph $G$ across the $s$ sites, we can assign each server's copy to a distinct edge $(u',v')\in S_u\times S_v$ in $G'$. See Fig. \ref{fig:lb-exam} for an illustrating example of the construction. Then we apply an algorithm for the without duplication model, \emph{e.g.}, the algorithm in Theorem \ref{thm:spanner-bl}, to get a $(2k-1)$-spanner $H'$ of $G'$. Finally, the coordinator computes a $(2k-1)$-spanner $H$ of $G$ by including an edge $(u,v)$ in $H$ if there is at least one edge between $S_u$ and $S_v$ in $H'$.

To show the constructed $H$ is a $(2k-1)$-spanner of $G$, let us consider an edge $(u,v)\in G$. By construction, there must be an edge $(u',v')\in S_u\times S_v$ in $G'$. Because $H'$ is a $(2k-1)$-spanner of $G'$, it contains a path $P'$ of length at most $(2k-1)\cdot W(u,v)$ between $u'$ and $v'$. For every edge $(x',y')$ in $P'$ where $x'\in S_x,y'\in S_y$, we have included an edge $(x,y)$ in $H$. Therefore, there exists a path $P$ of length at most $(2k-1)\cdot W(u,v)$ between $u$ and $v$ in $H$ and thus $H$ is a $(2k-1)$-spanner of $G$. Since the lower bound in the duplication model is $\Omega(s+n^{1+1/k}\log s)$ (Theorem \ref{thm:spanner-lb}), we have that the lower bound for the non-duplication model is $\Omega(s+(n/\sqrt{s})^{1+1/k}\log s)=\Omega(s+n^{1+1/k}s^{-1/2-1/(2k)}\log s)$. 

Since representing the result itself needs $\Omega(n^{1+1/k})$, combining this with the above result get the final lower bound, $\Omega(s+n^{1+1/k}\max\{1,s^{-1/2-1/(2k)}\log s\})$.
\qed

{\noindent \bf Discussions.}
%For constructing $(2k-1)$-spanners for $k>1$, we show an upper bound $\tilde{O}(s+n^{1+1/k})$, which is optimal (up to poly-logarithmic factors) in the with duplication model. We also prove a tighter lower bound $\Omega(s+s^{-1/2-1/(2k)}n^{1+1/k})$ in the without duplication model. Furthermore, we provide an algorithm for constructing $+2$-spanners or $3$-spanners using communication cost $\tilde{O}(s+n\sqrt{n+s})$, which is close to the lower bound $\Omega(s+n^{3/2})$ under the with duplication model. We observed that all developed algorithms in the blackboard model have significantly smaller communication costs than those in the message passing model. For instance, constructing $+2$ or $3$-spanners in the blackboard with duplication model incurs $\tilde{O}(s+n\sqrt{n+s})$ communication, much lower than $\tilde{O}(sn^{3/2})$ in the message passing model.
We highlight several interesting observations from our results in Table \ref{tab:distributed} and prior results in Table \ref{tab:distributed-mp}.
\begin{enumerate}
    % \item The communication cost in the blackboard without duplication model can be \emph{smaller} than that in the with duplication model because the allowed edge duplicates can incur extra communications. For example in $(2k-1)$-spanners, the upper bound $\tilde{O}(ks^{1-2/k}n^{1+1/k}+snk)$ in the without duplication model is smaller than $\tilde{O}(sn^{1+1/k})$ in the with duplication model.
    \item We demonstrate that for graph spanner constructions, the blackboard model is powerful to significantly reduce the communication complexity compared to the message passing model. For instance in duplication models, computing the $(2k-1)$-spanners incurs communication cost $\tilde{O}(sn^{1+1/k})$ in the message passing model but only $\tilde{O}(s+n^{1+1/k})$ in the blackboard model. This is not necessarily the case for all computing problems. For example, for computing the sum of bit vectors modulo two \cite{PVZ16} and estimating large moments \cite{WZ12}, the complexities are the same in both communication models.
    \item To trade better communication bounds, spanners constructed in a distributed manner may include more edges than the smallest number of edges required in a centralized model. For example in $+2$-spanners and $3$-spanners, the number of edges in the constructed structure is $n\sqrt{n+s}$, which is slightly larger than the optimal size $n\sqrt{n}$ in a centralized model. It is still open to investigate how to reduce the communication cost while maintaining an optimal number of edges in the spanner.
    \item For constructing $(2k-1)$-spanners, the upper bound $\tilde{O}(s+n^{1+1/k})$ with a logarithmic factor hidden is very close to the lower bound $\Omega(s+n^{1+1/k}\log s)$. There is a small gap between the upper bound $\tilde{O}(s+n\sqrt{n+s})$ and lower bound $\Omega(s+n^{3/2}\log s)$ for $+2$ or $3$-spanners. The gap is larger in $+k$-spanners (for $k>2$) where the lower bound becomes $\Omega(s+n^{4/3-o(1)}\log s)$. But this problem also happens in the message passing model. The construction of $+k$-spanners often involves more complex operations and might not be easy to adapt to distributed models.
\end{enumerate}

%Consider the pair-wise case of only preserving distances for some pairs of vertices, instead of all pairs, which is well-motivated in real-world applications. The PI has prior experience in developing pair-wise and source-wise spanners \cite{ZL17,ZL18,ZLB21a} and will further study distributed constructions of pair-wise spanners in the blackboard model to reduce the communication complexity in Table \ref{tab:distributed} under the new setting.

%In the \emph{without duplication} model, the upper bound $\tilde{O}(s+\sqrt{s/k}n^{3/2})$ is achieved by an adaptation of Algorithm 2 in \cite{FWY20} for the message passing model.

\section{Conclusions and Future Work}

In this paper, we propose the first set of algorithms that can perform distributed graph clustering and spectral sparsification under edge duplication in the two well-established communication models, the message passing and the blackboard models. We show the optimality of the achieved communication costs while maintaining a clustering quality nearly as good as a naive centralized method. We also perform the first investigation of distributed algorithms for constructing graph spanners in the blackboard under both duplication and non-duplication models.

As the future work, we will study how to achieve the optimal communication complexity for distributed graph clustering while relaxing the assumption made. Furthermore, most of the existing work concentrate on global clustering but ignore local clustering which only returns the cluster of a given seed vertex. We will devise a local clustering method that hopefully enjoys communication cost not dependent on the size of the input graph and is more communication-efficient than traditional global graph clustering methods.

Cut sparsifiers are another type of graph sparsifiers and they can approximately preserve all the graph cut values in the original graph. Although spectral sparsifiers are also cut sparsifiers, the latter might have smaller number of edges. Because the algorithm of \cite{KX16} can be generalized to cut sparsifiers, it is promising to adapt the techniques in this work to the new problem. Finally, it is an intriguing open problem to improve the upper bounds or lower bounds and close their gap in both duplication and non-duplication models.

\bibliographystyle{alpha}
% \small
\bibliography{dup}

\newcommand{\etalchar}[1]{$^{#1}$}
\begin{thebibliography}{CHKPY18}

\bibitem[AB16]{AB16}
A.~Abboud and G.~Bodwin.
\newblock {The 4/3 additive spanner exponent is tight}.
\newblock In {\em Proceedings of ACM STOC Conference}, pages 351--361, 2016.

\bibitem[ABS{\etalchar{+}}20]{ahmed2020graph}
Reyan Ahmed, Greg Bodwin, Faryad~Darabi Sahneh, Keaton Hamm, Mohammad
  Javad~Latifi Jebelli, Stephen Kobourov, and Richard Spence.
\newblock Graph spanners: A tutorial review.
\newblock {\em Computer Science Review}, 37:100253, 2020.

\bibitem[ADD{\etalchar{+}}93]{ADD+93}
I.~Althofer, G.~Das, D.P. Dobkin, D.~Joseph, and J.~Soares.
\newblock {On sparse spanners of weighted graphs}.
\newblock {\em Discrete Computational Geometry}, 9:81--100, 1993.

\bibitem[BKMP10]{BKM+10}
S.~Baswana, T.~Kavitha, K.~Mehlhorn, and S.~Pettie.
\newblock {Additive spanners and $(\alpha,\beta)$-spanners}.
\newblock {\em ACM Transactions on Algorithms}, 7(1), 2010.

\bibitem[BO15]{BO15}
Mark Braverman and Rotem Oshman.
\newblock The communication complexity of number-in-hand set disjointness with
  no promise.
\newblock In {\em Electron. Colloquium Comput. Complex.}, volume~22, page~2,
  2015.

\bibitem[Che13]{Chechik13}
S.~Chechik.
\newblock {New additive spanners}.
\newblock In {\em Proceedings of SIAM SODA Conference}, pages 498--512, 2013.

\bibitem[CHKPY18]{CKP+18}
K.~Censor-Hillel, T.~Kavitha, A.~Paz, and A.~Yehudayoff.
\newblock {Distributed construction of purely additive spanners}.
\newblock {\em Distributed Computing}, 31(3):223--240, 2018.

\bibitem[CSWZ16]{CSW+16}
J.~Chen, H.~Sun, D.P. Woodruff, and Q.~Zhang.
\newblock {Communication-optimal distributed clustering}.
\newblock In {\em Proceedings of NIPS Conference}, pages 3720--3728, 2016.

\bibitem[DOR21]{dershowitz2021communication}
Nachum Dershowitz, Rotem Oshman, and Tal Roth.
\newblock The communication complexity of multiparty set disjointness under
  product distributions.
\newblock In {\em Proceedings of ACM STOC Conference}, pages 1194--1207, 2021.

\bibitem[DS00]{DS00}
P.G. Doyle and J.L. Snell.
\newblock {Random walks and electric networks}.
\newblock {\em https://arxiv.org/abs/math/0001057}, 2000.

\bibitem[Erd64]{Erdos64}
P.~Erdos.
\newblock {Extremal problems in graph theory}.
\newblock {\em Theory of Graphs and Its Applications}, pages 29--36, 1964.

\bibitem[FHHP11]{FHH+11}
W.S. Fung, R.~Hariharan, N.~J.A. Harvey, and D.~Panigrahi.
\newblock {A general framework for graph sparsification}.
\newblock In {\em Proceedings of ACM STOC Conference}, pages 71--80, 2011.

\bibitem[FWY20]{FWY20}
M.V. Fernandez, D.P. Woodruff, and T.~Yasuda.
\newblock {Graph spanners in the message-passing model}.
\newblock In {\em Proceedings of ITCS Conference}, 2020.

\bibitem[Har12]{Har12}
N.~Harvey.
\newblock {Matrix concentration and sparsification}.
\newblock In {\em Workshop on Randomized Numerical Linear Algebra: Theory and
  Practise}, 2012.

\bibitem[KX16]{KX16}
I.~Koutis and S.C. Xu.
\newblock {Simple parallel and distributed algorithms for spectral graph
  sparsification}.
\newblock {\em ACM Transactions on Parallel Computing}, 3(2):14, 2016.

\bibitem[LGT14]{LGT14}
J.R. Lee, S.O. Gharan, and L.~Trevisan.
\newblock {Multiway spectral partitioning and higher-order Cheeger
  inequalities}.
\newblock {\em Journal of the ACM}, 61(6):37, 2014.

\bibitem[NJW01]{NJW01}
A.Y. Ng, M.I. Jordan, and Y.~Weiss.
\newblock {On spectral clustering: analysis and an algorithm}.
\newblock In {\em Proceedings of NIPS Conference}, pages 849--856, 2001.

\bibitem[PSZ15]{PSZ15}
R.~Peng, H.~Sun, and L.~Zanetti.
\newblock {Partitioning well-clustered graphs: spectral clustering works!}
\newblock In {\em Proceedings of COLT Conference}, pages 1423--1455, 2015.

\bibitem[PVZ16]{PVZ16}
J.M. Phillips, E.~Verbin, and Q.~Zhang.
\newblock {Lower bounds for number-in-hand multiparty communication complexity,
  made easy.}
\newblock {\em SIAM Journal on Computing}, 45(1):174--196, 2016.

\bibitem[SS11]{SS11}
D.A. Spielman and N.~Srivastava.
\newblock {Graph sparsification by effective resistances}.
\newblock {\em SIAM Journal on Computing}, 40(6):1913--1926, 2011.

\bibitem[ST11]{ST11}
D.A. Spielman and Shang-Hua Teng.
\newblock {Spectral sparsification of graphs}.
\newblock {\em SIAM Journal on Computing}, 40(4):981--1025, 2011.

\bibitem[SZ19]{SZ19}
H.~Sun and L.~Zanetti.
\newblock {Distributed graph clustering and sparsification}.
\newblock {\em ACM Transactions on Parallel Computing}, 6(3):17, 2019.

\bibitem[Tro12]{Tro12}
J.A. Tropp.
\newblock {User-friendly tail bounds for sums of random matrices}.
\newblock {\em Foundations of Computational Mathematics}, 12(4):389--434, 2012.

\bibitem[TZ05]{TZ05}
M.~Thorup and U.~Zwick.
\newblock {Approximate distance oracles}.
\newblock {\em Journal of the ACM}, 52(1):1--24, 2005.

\bibitem[VWW20]{vempala2020communication}
Santosh~S Vempala, Ruosong Wang, and David~P Woodruff.
\newblock The communication complexity of optimization.
\newblock In {\em Proceedings of SIAM SODA Conference}, pages 1733--1752. SIAM,
  2020.

\bibitem[WZ12]{WZ12}
D.P. Woodruff and Q.~Zhang.
\newblock {Tight bounds for distributed functional monitoring}.
\newblock In {\em Proceedings of ACM STOC Conference}, pages 941--960, 2012.

\bibitem[YX15]{YX15}
W.~Yang and H.~Xu.
\newblock {A divide and conquer framework for distributed graph clustering}.
\newblock In {\em Proceedings of ICML Conference}, pages 504--513, 2015.

\bibitem[ZLB21]{ZLB21a}
C.~Zhu, Q.~Liu, and J.~Bi.
\newblock {Spectral vertex sparsifiers and pair-wise spanners over distributed
  graphs}.
\newblock In {\em Proceedings of ICML Conference}, pages 12890--12900, 2021.

\bibitem[ZZL{\etalchar{+}}19]{ZZL+19a}
C.~Zhu, T.~Zhu, K.-Y. Lam, S.~Han, and J.~Bi.
\newblock {Communication-optimal distributed dynamic graph clustering}.
\newblock In {\em Proceedings of AAAI Conference}, pages 5957--5964, 2019.

\end{thebibliography}

\appendix

\section{Distributed Graph Clustering}
\label{adx:ss}

In this section, we will provide the missing proof of several theorems for distributed graph clustering, including Theorems \ref{thm:spanner-bl}, \ref{thm:distributed}, and \ref{thm:clustering}. 

\subsection{Proof of Theorem \ref{thm:spanner-bl}}
\setcounter{theorem}{1}
\begin{theorem}
The communication complexity of constructing a $(4k-2)$-spanner in weighted graphs under the blackboard with or without duplication model is $\tilde{O}(s+n^{1+1/k})$. In unweighted graphs, the stretch factor can be reduced to $2k-1$.
\end{theorem}

\proof
Considering weighted graphs, we first prove that the stretch factor of the structure $H$ output by Alg. \ref{alg:spanner} (after adaptations described in the main text) is $4k-2$. For each edge $(u, v) \in E$, if $(u, v) \not\in H$, it must be that including the edge $(u, v)$ would close a cycle of length $\leq 2k$. That is, there exists a path $P$ of $\leq 2k-1$ edges between $u$ and $v$ in $H$. Since we process edges in logarithmic scale, the edge weights in $P$ cannot be larger than $2W(u,v)$. Thus the path length of $P$ is at most $(4k-2)W(e)$. Therefore, the output $H$ is a $(4k-2)$-spanner. 

We then prove the communication cost. By construction, the output graph $H$ has girth (the minimum number of edges in a cycle contained in the graph) larger than $2k$. It is well known that a graph with girth larger than $2k$ have $O(n^{1+1/k})$ edges \cite{ADD+93}. Then $H$ always has $O(n^{1+1/k})$ edges throughout the processing of all logarithmic intervals. In addition, each site, if it does not modify $H$, needs to transmit a special marker of one bit to indicate that it has completed the processing. Therefore, the total communication cost is $\tilde{O}(s+n^{1+1/k})$. The algorithm works for both with and without duplication settings.

Finally, we prove the properties in the setting of unweighted graphs. The algorithm is provided in Alg. \ref{alg:spanner-unweighted}. By construction, for every edge $(u,v)$, $d_H(u,v)\leq (2k-1)d_G(u,v)$. Then the stretch factor is $2k-1$.

\begin{algorithm}[t]
\small
\caption{$(2k-1)$-spanners in the blackboard model with duplication model}
\renewcommand{\algorithmicrequire}{\textbf{Input:}}
\renewcommand{\algorithmicensure}{\textbf{Output:}}
\begin{algorithmic}[1]
\REQUIRE Unweighted graph $G(V,E)$ and a parameter $k>1$
\ENSURE Spanner $H$
\STATE $H\leftarrow \emptyset$\;
\FOR{each site $S_i$} 
    \FOR{each edge $e=(u,v)\in E_i$} 
        \IF{$d_H(u,v)>(2k-1)d_G(u,v)$}
            \STATE $H=H\cup \{e\}$\;
        \ENDIF
    \ENDFOR
    \IF{$H$ is updated in the above for-loop}
        \STATE Transmit the updated $H$ to the blackboard\;
    \ELSE
        \STATE Transmit a special marker to the blackboard to indicate the completion of the processing\;
    \ENDIF
\ENDFOR
\RETURN $H$;
\end{algorithmic}
\label{alg:spanner-unweighted}
\end{algorithm}

For communication cost, the output graph has girth larger than $2k$ by construction. Furthermore, each site needs to transmit a special marker of one bit when it does not update the current spanner $H$. Therefore, the total communication cost is $\tilde{O}(s+n^{1+1/k})$. 
\qed

\subsection{Proof of Theorem \ref{thm:distributed}}
We start by defining a few notations.
Suppose that $P$ is a path connecting the two endpoints of an edge $e$, the \emph{stretch} of $e$ over $P$ is equal to $\alpha_P(e)=W(e)\sum_{e'\in P}(1/W(e'))$.
The notation $L_{A}\preceq L_{B}$ means that for every vector $x\in R^n$, $x^TL_Ax\leq x^TL_Bx$, while $L_{A}\preceq^{\{0,1\}} L_{B}$ means that for every vector $x\in \{0,1\}^n$, $x^TL_Ax\leq x^TL_Bx$. 
The Laplacian matrix $L_G^e$ of an edge $e$ in $G$ is the Laplacian matrix of the subgraph of $G$ containing only the edge $e$.
It is zero elsewhere except a $2\times 2$ submatrix.
We will also use the following variant \cite{Har12} of a matrix concentration bound by \cite{Tro12}.

\setcounter{theorem}{6}
\begin{theorem}
\label{thm:mcb}
\cite{Har12} Let $Y_1,\cdots,Y_k$ be independent positive semi-definite matrices of size $n\times n$.
Let $Y=\sum_{i=1}^kY_i$ and $Z=E[Y]$.
Suppose for every $i\in [1,k]$, $Y_i\preceq SZ$, where $S$ is a scalar.
Then for all $\epsilon\in [0,1]$,
$Pr[\sum_{i=1}^kY_i\preceq (1-\epsilon)Z]\leq n\cdot exp(-\epsilon^2/2S)$, and $Pr[\sum_{i=1}^kY_i\succeq (1+\epsilon)Z]\leq n\cdot exp(-\epsilon^2/3S)$.
\end{theorem}

\setcounter{theorem}{2}
\begin{theorem}[Spectral Sparsification under Edge Duplication]
For a distributed graph $G$ and parameters $\epsilon\in (0,1)$ and $\rho=O(\log n)$, Alg. \ref{alg:ss} can construct a $(1+\epsilon)$-spectral sparsifier for $G$ of expected size $\tilde{O}(n)$ using communication cost $\tilde{O}(ns)$ and $\tilde{O}(n+s)$ in the message passing and blackboard with duplication models respectively, with probability at least $1-n^{-c}$ for constant $c$.
\end{theorem}

\proof
The communications happen for logarithmic iterations of distributed spanner constructions and during transmitting sampled edges in the last iteration. 
Constructing a $t$-bundle $\log n$-spanner involves computation of a $\log n$-spanner for $t=O(\epsilon^{-2}\log n)$ times, thereby incurring communication cost of $\tilde{O}(n+s)$ and $\tilde{O}(ns)$ under the message passing and blackboard models, respectively.
As we will prove shortly, the output sparsifier has size $\tilde{O}(n)$.
Because the sampled edges in the last iteration are a part of the output, It also has size $\tilde{O}(n)$.
Therefore, the total communication costs are $\tilde{O}(n+s)$ and $\tilde{O}(ns)$, respectively.

It is easy to see by simple mathematical calculations that the probability of sampling a non-spanner edge across sites is $0.25$.
Then the proof that the output is a spectral sparsifier of the input graph follows directly from \cite{KX16}. For self-containedness, we provide the proof below. 

We first prove that $R_{G}(e)\leq \log n/t\cdot W(e)$ and $W(e)\cdot L_{G}^e\preceq \log n/t\cdot L_{G}$ for $t=O(\epsilon^{-2}\log n)$. By construction, for every edge $e\in G-J$, there are $t$ edge-disjoint paths $P_1,\cdots,P_t$ between the two endpoints of $e$ in $J$, such that for every $i\in [1,t]$, $\alpha_{P_i}(e)\leq \log n$.
By definition, for every $i\in [1,t]$, we have that
\begin{equation}
\label{eq:alpha}
\alpha_{P_i}(e)=W(e)\sum_{e\in P_i}(1/W(e))\leq \log n.
\end{equation}
According to the formula for resistors connected in series, for every path $P_i$ with $i\in[1,t]$, the effective resistance between the two endpoints of $e$ in $P_i$ is equal to
\begin{equation}
\label{eq:serial}
R_{P_i}(e)=\sum_{e\in P_i}R(e)=\sum_{e\in P_i}(1/W(e))
\end{equation}
Combining Equations (\ref{eq:alpha}) and (\ref{eq:serial}), we have that for every $i\in [1,t]$, $R_{P_i}(e)\leq \log n/W(e)$.
According to the formula for resistors connected in parallel, for a set of edge-disjoint paths $\{P_1,\cdots,P_t\}$ between $e$'s two endpoints, and let $P$ be the union of these paths $P=\cup_{i=1}^tP_i$, the effective resistance between $e$'s two endpoints in $P$ is equal to
$R_P(e)=(\sum_{i=1}^t(R_{P_i}(e))^{-1})^{-1} \leq \log n/t\cdot W(e).$
According to the Rayleigh's monotonicity law \cite{DS00}, for any subgraph $H$ of $G$ and any edge $e\in G$, $R_G(e)\leq R_H(e)$ holds.
Therefore,
\begin{equation}
\label{eq:resistance}
R_{G}(e)\leq R_P(e)\leq \log n/t\cdot W(e).
\end{equation}
By \cite{SS11}, we have 
\begin{equation}
\label{eq:Lewf}
L_G^e\preceq R_G(e)L_G.
\end{equation}
By combining Equation (\ref{eq:Lewf}) with Equation (\ref{eq:resistance}), we have that
\begin{equation}
\label{eq:resistance1}
W(e)\cdot L_{G}^e\preceq \log n/t\cdot L_{G}.
\end{equation}

Next, we prove that the output $H$ of \emph{Light-SS} is a $(1+\epsilon)$-spectral sparsifier.
For every edge $e\in G-J$, let $X_e$ be the random variable defined as %$4W(e)L_{G-F}^e$ with probability $0.25$ and 0 with probability $0.75$.

\begin{equation*}
    X_e=
    \begin{cases}
      4W(e)L_{G}^e, & \text{with probability}\ 0.25 \\
      0, & \text{otherwise}
    \end{cases}
\end{equation*}

For every $i\in [1,(\lfloor \epsilon^2/(6\log n) \rfloor)^{-1}]$, let $J_i=\lfloor \epsilon^2/(6\log n) \rfloor J$, which implies that
\begin{equation*}
L_{J_i}=\lfloor \epsilon^2/(6\log n) \rfloor L_{J}.
\end{equation*}
We then apply Theorem \ref{thm:mcb} to the random matrix
\begin{equation*}
\begin{aligned}
Y&=\sum_{e\in G-J}X_e+\sum_{i=1}^{(\lfloor \epsilon^2/(6\log n) \rfloor)^{-1}}L_{J_i}=\sum_{e\in G-J}X_e+L_{J}.
\end{aligned}
\end{equation*}
Note that
\begin{equation*}
\begin{aligned}
E(Y)&=E(\sum_{e\in G-J}X_e+L_{J})=\sum_{e\in G-J}E(X_e)+L_{J}\\&=\sum_{e\in G-J}L_{G}^e+L_{J}=L_{G}.
\end{aligned}
\end{equation*}
By the definition of $X(e)$ and Equation (\ref{eq:resistance1}), for every $e\in G-J$ we have that
\begin{equation*}
X(e)\preceq 4W(e)\cdot L_{G}^e\preceq \epsilon^2/(6\log n)\cdot L_{G}.
\end{equation*}
Furthermore, by definition of $J_i$ and the fact that $L_{J}\preceq L_{G}$, we have for every $i\in [1,(\lfloor \epsilon^2/(6\log n) \rfloor)^{-1}]$,
\begin{equation*}
L_{J_i}=\lfloor \epsilon^2/(6\log n) \rfloor\cdot L_{J}\preceq \epsilon^2/(6\log n)\cdot L_{G}.
\end{equation*}
Now the condition of Theorem \ref{thm:mcb} is satisfied with $S=\epsilon^2/(6\log n)$.
Therefore, the inequality
\begin{equation}
\label{eq:ftss}
(1-\epsilon)L_{G}\preceq L_{H}\preceq (1+\epsilon)L_{G}
\end{equation}
holds with probability at least $1-1/2n\cdot exp(-3\log n)=1-1/2n^{-2}$.

Finally, we prove that the property of the main algorithm, Alg. \ref{alg:ss}. By the property of \emph{Light-SS} proved above and Induction, for every $i\in [1,\lceil\log\rho\rceil]$, the event that the inequality
\begin{equation*}
(1-\epsilon/\lceil \log \rho \rceil)^iL_{G}\preceq L_{G_i}\preceq (1+\epsilon/\lceil \log \rho \rceil)^iL_{G}
\end{equation*}
holds and the expected size is
\begin{equation*}
O(ni\log^2n\log^2\rho/\epsilon^2+m/2^i),
\end{equation*}
happens with probability at least $(1-1/n^2)^i$.
Since \emph{SS} outputs $H=G_{\lceil \log \rho \rceil}$ as the final spectral sparsifier, the expected size becomes $O(n\log^3n\log^3\rho/\epsilon^2+m/\rho)$. Because $\epsilon\in(0,1)$ and $\rho=O(\log n)$, the expected size is $\tilde{O}(n)$.
The desirable properties hold with probability at least $1-n^{-c}$ for constant $c$.
\qed

\subsection{Proof of Theorem \ref{thm:clustering}}
For every node set $S\subseteq V$ in $G$, let its \emph{volume} and \emph{conductance} be $vol_G(S)=\sum_{u\in S,v\in V}W(u,v)$ and $\phi_G(S)=(\sum_{u\in S,v\in V-S}W(u,v))/vol_G(S)$, respectively.
Intuitively, a small value of conductance $\phi(S)$ implies that nodes in $S$ are likely to form a cluster.
%Intuitively, a small value of conductance $\phi(S)$ means that nodes in $S$ have few connections with the rest nodes in $V$, and thus implies that nodes in $S$ formulate a cluster.
A collection of subsets $A_1,\cdots,A_k$ of nodes is called a \emph{(k-way) partition} of $G$ if (1) $A_i\cap A_j=\emptyset$ for $1\leq i\not=j\leq k$; and (2) $\cup_{i=1}^kA_i=V$.
The \emph{k-way expansion constant} is defined as $\rho(k)=\min_{partition A_1,\cdots,A_k}\max_{i\in[1,k]}\phi(A_i)$.
Let the \emph{normalized Laplacian matrix} of $G$ be $\mathcal{L}_G=D_G^{-1/2}L_GD_G^{-1/2}$ and its eigenvalues are $\lambda_1(\mathcal{L}_G)\leq \cdots\leq \lambda_n(\mathcal{L}_G)$.
The high-order Cheeger inequality shows that $\lambda_k/2\leq \rho(k)\leq O(k^2)\sqrt{\lambda_k}$ \cite{LGT14}.
A lower bound on $\Upsilon_G(k)=\lambda_{k+1}/\rho(k)$ implies that, $G$ has exactly $k$ well-defined clusters \cite{PSZ15}.
It is because a large gap between $\lambda_{k+1}$ and $\rho(k)$ guarantees the existence of a k-way partition $A_1,\cdots,A_k$ with bounded $\phi(A_i)\leq \rho(k)$, and that any $(k+1)$-way partition $A_1,\cdots,A_{k+1}$ contains a subset $A_i$  with significantly higher conductance $\rho(k+1)\geq \lambda_{k+1}/2$ compared with $\rho(k)$.
For any two sets $X$ and $Y$, the symmetric difference of $X$ and $Y$ is defined as $X\Delta Y=(X-Y)\cup(Y-X)$.
To prove Theorem \ref{thm:clustering}, we will use the following lemma and theorems.

\begin{lemma}{\cite{CSW+16}}
\label{thm:conductance}
Let $H$ be a $(1+\epsilon)$-spectral sparsifier of $G(V,E)$ for some $\epsilon\leq 1/3$.
For all node sets $S\subseteq V$, the inequality $0.5\cdot\phi_G(S)\leq \phi_H(S)\leq 2\cdot\phi_G(S)$ holds.
\end{lemma}

\setcounter{theorem}{7}
\begin{theorem}
\label{thm:spectralclustering}
\cite{PSZ15} Given a graph $G$ with $\Upsilon_G(k)=\Omega(k^3)$ and an optimal partition $S_1,\cdots,S_k$ achieving $\rho(k)$ for some positive integer $k$, the spectral clustering algorithm can output partition $A_1,\cdots,A_k$ such that, for every $i\in [1,k]$, the inequality $vol(A_i\Delta S_i)=O(k^3\Upsilon^{-1}vol(S_i))$ holds.
\end{theorem}

\setcounter{theorem}{3}
\begin{theorem}
For a distributed graph $G$ with $\Upsilon_G(k)=\Omega(k^3)$ and an optimal partition $P_1,\cdots,P_k$ achieving $\rho(k)$ for some positive integer $k$, there exists an algorithm that can output partition $A_1,\cdots,A_k$ at the coordinator such that for every $i\in [1,k]$, $vol(A_i\Delta P_i)=O(k^3\Upsilon^{-1}vol(P_i))$ holds with probability at least $1-n^{-c}$ for constant $c$. The communication costs in the message passing and blackboard with duplication models are $\tilde{O}(ns)$ and $\tilde{O}(n+s)$, respectively.
\end{theorem}

\proof
The algorithm starts by distributively constructing a spectral sparsifier $H$ of $G$ in the coordinator using our algorithms in Theorem \ref{thm:distributed} and then applies a standard graph clustering algorithm, \emph{e.g.}, spectral clustering \cite{NJW01} in $H$ to get the clustering results. The communication costs directly follow from Theorem \ref{thm:distributed} since the final clustering step does not incur communications. The rest of the proof follows from \cite{CSW+16} and we present it for the sake of self-containedness.

We prove that if $G$ satisfies that $\Upsilon_{G}(k)=\Omega(k^3)$, $H$ also satisfies that $\Upsilon_{H}(k)=\Omega(k^3)$.
By the definition of $\Upsilon$, it suffices to prove that $\rho_{H}(k)=\Theta(\rho_{H}(k))$ and $\lambda_{k+1}(\mathcal{L}_{H})=\Theta(\lambda_{k+1}(\mathcal{L}_{G}))$.
The former follows from that for every $i\in [1,k]$, the inequality 
\begin{equation*}
0.5\cdot\phi_{G}(S_i)\leq \phi_{H}(S_i)\leq 2\cdot\phi_{G}(S_i)
\end{equation*}
holds, according to Lemma \ref{thm:conductance}.
According to the definition of $(1+\epsilon)$-spectral sparsifier and simple math, it holds for every vector $x\in R^n$ that
\begin{equation*}
\begin{aligned}
(1-\epsilon)x^TD_{G}^{-1/2}L_{G}D_{G}^{-1/2}x & \leq x^TD_{G}^{-1/2}L_{H}D_{G}^{-1/2}x \\
& \leq (1+\epsilon)x^TD_{G}^{-1/2}L_{G}D_{G}^{-1/2}x.
\end{aligned}
\end{equation*}
By the definition of normalized graph Laplacian $\mathcal{L}_G$, and the fact that for every vector $y\in R^n$,
\begin{equation*}
0.5\cdot y^TD_{G}^{-1}y\leq y^TD_{H}^{-1}y\leq 2y^TD_{G}^{-1}y,
\end{equation*}
we have that for every $i\in [1,n]$, 
\begin{equation*}
\lambda_i(\mathcal{L}_{H})=\Theta(\lambda_i(\mathcal{L}_{G})),
\end{equation*}
which implies that $\lambda_{k+1}(\mathcal{L}_{H})=\Theta(\lambda_{k+1}(\mathcal{L}_{G}))$.
Then we can apply the spectral clustering algorithm in $H$ to get the desirable properties, according to Theorem \ref{thm:spectralclustering}.
\qed

\section{Graph Spanners}

\subsection{BFS in the Blackboard Model}
\label{adx:bfs}

\begin{algorithm}[t]
\small
\caption{\emph{BFS} in the blackboard model with duplication model}
\renewcommand{\algorithmicrequire}{\textbf{Input:}}
\renewcommand{\algorithmicensure}{\textbf{Output:}}
\begin{algorithmic}[1]
\REQUIRE Graph $G(V,E)$ and root vertex $u\in V$
\ENSURE BFS tree $T$
\STATE $T\leftarrow \{u\}$; $A\leftarrow \{u\}$; $C\leftarrow \emptyset$\;
\WHILE{$C\not=V$} 
    \FOR{each site $S_i$} 
        \STATE $S_i$ transmits its edges $(v,w)\in E_i$ such that $v\in A$ and $w\not\in C$ and $(v,w)$ was not transmitted by sites $S_j$ for $j<i$\; \label{bfs:edge}
        \STATE If such an edge cannot be found, $S_i$ transmits a special marker to indicate the completion of its processing\;
        \STATE The coordinator includes the received edges $(v,w)$ into $T$ and maintains $N=\{w\,|\,(v,w)\}$\;
    \ENDFOR
    \STATE $C\leftarrow C \cup A$;  $A\leftarrow N$\;
\ENDWHILE
\RETURN $T$;
\end{algorithmic}
\label{alg:bfs}
\end{algorithm}

Here we discuss an important building block for graph spanner construction: growing a breath first search (BFS) tree from a root vertex in a distributed graph.
We observe that the communication complexity of computing a BFS tree from a given vertex in the blackboard model with or without duplication is $\tilde{O}(n+s)$, which is significantly smaller than $\tilde{O}(ns)$ in the message passing model \cite{FWY20}. This can be achieved by a simple distributed protocol where the blackboard maintains a partial BFS tree and an active set $A$ of vertices, both initialized to be the root vertex. In each iteration, each of the sites $S_i$ in order transmits edges of vertices in $A$ pointing to a vertex that has never been in the active set and has not transmitted previously by site $S_j$ for $j<i$. If such an edge cannot be found, $S_i$ submits a special marker to indicate the completion of its processing. At the end of each iteration, the current active set $A$ is updated to be the other endpoints of the transmitted edges. Since the edge linking each vertex to the BFS tree is transmitted at most once and each site has to transmit a special marker if it cannot add a new edge, the incurred communication cost is $\tilde{O}(n+s)$. The formal pseudo-code can be found in Alg. \ref{alg:bfs}.

\setcounter{theorem}{8}
\begin{theorem}
\label{thm:bfs}
The communication complexity of building a BFS tree in the blackboard with or without duplication model is $\tilde{O}(n+s)$.
\end{theorem}

\proof
We prove it for the more general duplication model. In Alg. \ref{alg:bfs}, the edge linking each vertex to the BFS tree is transmitted at most \emph{once}. This is because in Line \ref{bfs:edge}, site $S_i$ would not transmit an edge $(v,w)$ if the edge is already transmitted previously by some site $S_j$ for $j<i$. This is possible in the blackboard model since edges sent to the blackboard by one site are visible to all other sites. Furthermore, each site has to transmit one bit of information if it does not send any edge. Therefore, the total communication cost is $\tilde{O}(n+s)$.
\qed

\subsection{$+2$-spanners and $3$-spanners}

\begin{algorithm}[t]
\small
\caption{$+2$-spanners in the blackboard without duplication model}
\renewcommand{\algorithmicrequire}{\textbf{Input:}}
\renewcommand{\algorithmicensure}{\textbf{Output:}}
\begin{algorithmic}[1]
\REQUIRE Graph $G(V,E)$
\ENSURE Spanner $H$
\STATE $H\leftarrow \emptyset$\;
\FOR{each site $S_i$} %\COMMENT{Include all edges of vertices with degree at most $\sqrt{n+s}$ in $H$}
    \FOR{each vertex $u$}
        \IF{the sum of the number of $u$'s edges in the blackboard and the number of $u$'s edges in $E_i$ is no larger than $\sqrt{n+s}$}
            \STATE $S_i$ transmits all its edges $(u,v)\in E_i$ and then the coordinator includes these edges into $H$\;
        \ELSE
            \STATE $S_i$ transmits a special marker to inform the completion of its processing\;
        \ENDIF
    \ENDFOR
\ENDFOR

\STATE The coordinator samples $\tilde{O}(n/\sqrt{n+s})$ vertices $R$ uniformly at random with replacement from all the vertices $V$\; %\COMMENT{Growing BFS trees}
\FOR{each sampled vertex $v\in R$}
    \STATE The coordinator grows a BFS tree from $v$ and includes edges in the BFS tree into $H$\;
\ENDFOR
\RETURN $H$;
\end{algorithmic}
\label{alg:+2}
\end{algorithm}

{\noindent \bf Upper Bound.}
We show that the communication complexity of constructing $+2$-spanners in the blackboard \emph{without} edge duplication is $\tilde{O}(s+n\sqrt{n+s})$ (Theorem \ref{thm:+2u}). It is achieved by a simple distributed algorithm as provided in Alg. \ref{alg:+2}. First, we aim to include all the edges of vertices with degree at most $\sqrt{n+s}$ in the spanner. However, the vertex degrees are not given directly. A naive method is that each site transmits all vertex degrees to the blackboard who then takes their sum. But this incurs very costly communication $\tilde{O}(ns)$. Our solution is that each site $S_i$ in order transmits each vertex $u$'s edges in its edge set $E_i$ if the sum of the number of $u$'s edges in the blackboard and the number of $u$'s edges in $E_i$ is no larger than $\sqrt{n+s}$. This only incurs communication cost $\tilde{O}(s+n\sqrt{n+s})$, instead of $\tilde{O}(ns)$. Next, the coordinator samples $\tilde{O}(n/\sqrt{n+s})$ vertices uniformly at random with replacement from all the vertices and let the sampled set be $R$. It then grows a BFS tree from each sampled vertex in $R$ using Alg. \ref{alg:bfs}, and includes edges of the BFS trees in the spanner. Our results are summarized in Theorem \ref{thm:+2u}.
%This requires communication cost of $\tilde{O}(n\sqrt{n+s})$ since growing a BFS tree incurs $\tilde{O}(n+s)$ communication. Therefore, the total communication cost is $\tilde{O}(s+n\sqrt{n+s})$. 

\begin{theorem}
\label{thm:+2u}
The communication complexity of constructing a $+2$-spanner in the blackboard with or without duplication model is $\tilde{O}(s+n\sqrt{n+s})$.
\end{theorem}

\proof
We first prove the distance surplus $+2$. Consider the collection $C$ of (immediate) neighbors of vertices of degree at least $\sqrt{n+s}$ in $G$. The event $X$ that the sample set $R$ contains at least one vertex from each set of neighbors in $C$ happens with probability at least $1-o(1)$. This can be obtained by a direct application of a well-known sampling fact, Lemma \ref{lemma:sample} with $U=V$ and $t=\sqrt{n+s}$.

\setcounter{theorem}{10}
\begin{lemma}[Lemma 8 in \cite{FWY20}]
\label{lemma:sample}
Let $C$ be a collection of sets over a ground set $U$ each of size at least $t$. If we sample $|U|/t\cdot \log|C/\delta|$ elements from $U$ uniformly with replacement, with probability at least $1-\delta$ we sample at least one element from each set in $C$.
\end{lemma}

Consider the shortest path $P$ between two vertices $u,v$ in $G$. If all edges on $P$ are present in $H$, then $d_H(u,v)=d_G(u,v)$ and the distance surplus trivially holds. Otherwise, let $(u',v')\in P$ be a missing edge in $H$. We know both $u'$ and $v'$ have degree at least $\sqrt{n+s}$ as otherwise all their edges are included in $H$. Suppose the event $X$ occurs and $x\in R$ is a neighbor of $u'$. Then we have

\begin{equation*}
\begin{aligned}
    d_H(u,v)&\leq d_H(u,x)+d_H(x,v)\\
    &=d_G(u,x)+d_G(x,v)\\
    &\leq d_G(u,u')+1+d_G(u',v)+1\\
    &= d_G(u,v)+2.
\end{aligned}
\end{equation*}

The first and third inequalities follow from the triangle inequality. The second equality holds since edges in the BFS tree rooted at $x$ are included in $H$. The last equality holds because $u'$ lies on the shortest path $P$.

We now prove the communication cost. Our method of including all edges of vertices with degree at most $\sqrt{n+s}$ incurs communication cost $\tilde{O}(s+n\sqrt{n+s})$. Constructing all the $\tilde{O}(n/\sqrt{n+s})$ BFS trees requires communication cost $\tilde{O}(n\sqrt{n+s})$ since growing each BFS tree incurs $\tilde{O}(n+s)$ communication according to Theorem \ref{thm:bfs}. Therefore, the total communication cost is $\tilde{O}(s+n\sqrt{n+s})$.
We point out that this method can be adapted to the duplication model. The modification is in the implementation of including all the edges of vertices with degree at most $\sqrt{n+s}$. When edge duplicates across sites are allowed, each site $S_i$ checks whether the number of $u$'s edges in the blackboard and the number of distinct edges associated with $u$ in $E_i$ of $S_i$ is no larger than $\sqrt{n+s}$. If so, it only transmits the distinct edges of $u$ to the blackboard, excluding any edges already in the blackboard. It is easy to see that both the correctness and the communication cost are not affected by this adaption.
\qed

Since the $+2$-spanner construction algorithm immediately gives a $3$-spanner construction algorithm in unweighted graphs, communication upper bounds of computing $3$-spanners follow from Theorem \ref{thm:+2u}, as shown in Corollary \ref{cor:3u}.

\begin{corollary}
\label{cor:3u}
The communication complexity of constructing a $3$-spanner in the blackboard with or without duplication model is $\tilde{O}(s+n\sqrt{n+s})$.
\end{corollary}

{\noindent \bf Lower Bound.}
We first consider the duplication model and then will extend to the non-duplication model.
For the duplication model, we can prove the communication lower bound of computing $+2$-spanners is $\Omega(s+n^{3/2}\log s)$. It is because the size lower bound of $+2$-spanners is $\Omega(n^{3/2}\log s)$ as well as Lemma \ref{lemma:lb}. There is only a small gap between the upper bound and the lower bound with an approximation ratio of $\sqrt{(n+s)/n}$. 

The lower bound $\Omega(s+n^{3/2})$ for constructing $3$-spanners in the duplication model can also be derived using Lemma \ref{lemma:lb}. See Theorem \ref{thm:+2l} for the formal presentation. Computing $3$-spanners is weaker than computing $+2$-spanners and thus might enjoy better lower bounds. However, how to achieve a tighter bound for $3$-spanners remain open in both the message passing and the blackboard models.

\begin{theorem}
\label{thm:+2l}
The communication complexity of constructing a $+2$-spanner or a $3$-spanner in the blackboard with duplication model is $\Omega(s+n^{3/2}\log s)$.
\end{theorem}

\proof
According to Lemma \ref{lemma:+2}, there is a graph $G$ on $n$ vertices with size $\Theta(n^{1.5})$ and girth at least $6$. Since removing any edge in $G$ increases the distance from its endpoints to at least $5$, the only $+2$-spanner ($3$-spanner, respectively) of $G$ is $G$ itself. Then by applying $f(n)=n^{1.5}$ in Lemma \ref{lemma:lb}, we get the desired lower bound $\Omega(s+n^{3/2}\log s)$.

\begin{lemma}[Lemma 5 in \cite{FWY20}]
\label{lemma:+2}
For every $n$, there is a family of graphs on $n$ vertices with $\Theta(n^{1.5})$ edges and girth at least $6$.
\end{lemma}
\qed

We now consider lower bounds in the non-duplication model, where we cannot use the technique in Lemma \ref{lemma:lb}. But, we provide a weaker and similar lemma, Lemma \ref{lemma:lb-non}, for the non-duplication model. By incorporating Lemma \ref{lemma:lb-non} into the analysis of Theorems \ref{thm:+2l}, we get the lower bound of computing $+2$-spanner or $3$-spanners, $\Omega(s+n^{3/2})$, as shown in Theorem \ref{thm:+2l-non}.

\begin{lemma}
\label{lemma:lb-non}
Suppose there exists an $n$-vertex graph $F$ of size $f(n)$ such that $F$ is the only spanner of itself or no proper subgraph $F'$ of $F$ is a spanner. Then the communication complexity of computing a spanner in the blackboard without duplication model is $\Omega(s+f(n))$ bits. 
\end{lemma}

\proof
First, representing the result itself needs $f(n)$ bits. Next, each site needs to transmit at least one bit of information to inform the completion of its processing if it does not transmit some edge(s). Therefore, the total communication cost of constructing a spanner is $\Omega(s+f(n))$ bits.
\qed

\begin{theorem}
\label{thm:+2l-non}
The communication complexity of constructing a $+2$-spanner or a $3$-spanner in the blackboard without duplication model is $\Omega(s+n^{3/2})$.
\end{theorem}

\subsection{$+k$-Spanners}
The communication upper bound of constructing $+k$-Spanners, $\tilde{O}(s+n\sqrt{n+s})$, immediately follows from the upper bound of constructing $+2$-spanners, because $+2$-spanners are valid $+k$-spanners.
In the duplication and non-duplication models, the communication lower bounds $\Omega(s+n^{4/3-o(1)}\log s)$ and $\Omega(s+n^{4/3-o(1)})$ can be obtained by using the size lower bound $\Omega(n^{4/3-o(1)})$ of $(+k)$-spanners \cite{AB16} and Lemma \ref{lemma:lb} and \ref{lemma:lb-non}, respectively.

\begin{theorem}
\label{thm:+kub}
The communication complexity of constructing a $+k$-spanner in the blackboard with or without duplication model is $\tilde{O}(s+n\sqrt{n+s})$.
\end{theorem}

\begin{theorem}
\label{thm:+kl}
The communication complexity of constructing a $+k$-spanner in the blackboard with duplication model is $\Omega(s+n^{4/3-o(1)}\log s)$.
\end{theorem}

\begin{theorem}
\label{thm:+kl-non}
The communication complexity of constructing a $+k$-spanner in the blackboard without duplication model is $\Omega(s+n^{4/3-o(1)})$.
\end{theorem}

\end{document}